\documentclass[a4paper,11pt]{article}
\pdfoutput=1 

\usepackage{jinstpub} 
\usepackage{graphicx}
\usepackage{subcaption}
\usepackage{mhchem}
\usepackage{amssymb}

\title{Irradiation and performance of RGB-HD Silicon Photomultipliers for calorimetric applications}

\newcommand{\nima}        [3]  {{\emph{Nucl.\ Instrum.\ Meth.\ A }}~{\bf #1} (#2) #3}
\newcommand{\itns}        [3]  {{\emph{IEEE Trans. Nucl. Sci.}}~{\bf #1} (#2) #3}
\newcommand{\jinst}       [3]  {{\emph{JINST}}~{\bf #1} (#2) #3}

\newcommand{\um}{\upmu\text{m}}
\newcommand{\ncmsq}{\text{n}/\text{cm}^2}
\newcommand{\kethree}{$K^+ \rightarrow e^+ \pi^0 \nu_e$}

\author[a]{F.~Acerbi,}
\author[b,c]{G.~Ballerini,}
\author[b,c]{A.~Berra,}
\author[b,c]{C.~Brizzolari,} 
\author[d]{G.~Brunetti,} 
\author[e]{M.G.~Catanesi,} 
\author[f]{S.~Cecchini,} 
\author[f]{F.~Cindolo,}
\author[c,g]{A.~Coffani,}
\author[d,h]{G.~Collazuol,}
\author[d]{E.~Conti,}
\author[d]{F.~Dal Corso,}
\author[c,g]{C.~Delogu,}
\author[i]{G.~De Rosa,}
\author[a]{A.~Gola,}
\author[e]{R.~A.~Intonti,}
\author[l,m]{C.~Jollet,}
\author[n]{Y.~Kudenko,}
\author[d,h]{A.~Longhin,}
\author[p]{L.~Ludovici,} 
\author[i]{L.~Magaletti,} 
\author[f]{G.~Mandrioli,}
\author[f]{A.~Margotti,}  
\author[b,c]{V.~Mascagna,}
\author[f,o]{N.~Mauri,}
\author[m]{A.~Meregaglia,}
\author[d,h]{M.~Pari,}
\author[f,o]{L.~Pasqualini,}
\author[a]{G.~Paternoster,}
\author[f]{L.~Patrizii,}
\author[a]{C.~Piemonte,} 
\author[f]{M.~Pozzato,} 
\author[d]{F.~Pupilli,}
\author[b,c]{M.~Prest,}
\author[e]{E.~Radicioni,}
\author[i]{C.~Riccio,}
\author[i]{A.C.~Ruggeri,}
\author[h]{C.~Scian,} 
\author[f]{G.~Sirri,} 
\author[b,c]{M.~Soldani,} 
\author[c,g]{M.~Tenti,} 
\author[c,g]{M.~Torti,} 
\author[c,g,1]{F.~Terranova, \note{Corresponding author.}} 
\author[q]{E.~Vallazza}

\affiliation[a]{Fondazione Bruno Kessler, Via Sommarive 18, Povo (TN), Italy}
\affiliation[b]{Universit\`a degli Studi dell'Insubria, Via Valleggio 11, Como, Italy}
\affiliation[c]{INFN Milano Bicocca, Piazza della Scienza 3, Milano, Italy}
\affiliation[d]{INFN Padova, Via Marzolo 8, Padova, Italy}
\affiliation[e]{INFN Sezione di Bari, Via E. Orabona 4, Bari, Italy}
\affiliation[f]{INFN Sezione di Bologna, Via Berti Pichat 6, Bologna, Italy}
\affiliation[g]{Universit\`a degli Studi di Milano Bicocca, Piazza della Scienza 3, Milano, Italy}
\affiliation[h]{Universit\`a degli Studi di Padova, Via Marzolo 8, Padova, Italy}
\affiliation[i]{INFN Napoli and Universit\`a degli Studi di Napoli, Via Cintia, Napoli, Italy}
\affiliation[l]{Institute Pluridisciplinaire Hubert Curien, 23 rue du Loess, Strasbourg, France}
\affiliation[m]{Centre de Etudes Nucleaires de Bordeaux Gradignan, 19 Chemin du Solarium, Bordeaux, France}
\affiliation[n]{Institute for Nuclear Research of the Russian Academy of Sciences, Moscow, Russia}
\affiliation[o]{Universit\`a degli Studi di Bologna, Via Irnerio 46, Bologna, Italy}
\affiliation[p]{INFN Roma, Piazzale Aldo Moro 2, Roma, Italy}
\affiliation[q]{INFN Trieste, Padriciano 99, 34012 Trieste, Italy}

\emailAdd{francesco.terranova@cern.ch}

\abstract{Silicon Photomultipliers with cell-pitch ranging from 12~$\um$ to
  20~$\um$ were tested against neutron irradiation at moderate
  fluences to study their performance for calorimetric applications. The photosensors were
  developed by FBK employing the RGB-HD technology. We performed
  irradiation tests up to $2 \times 10^{11} \ \ncmsq$ (1 MeV eq.)  at
  the INFN-LNL Irradiation Test facility. The SiPMs were characterized
  on-site (dark current and photoelectron response) during and after
  irradiations at different fluences. The irradiated SiPMs were
  installed in the ENUBET compact calorimetric modules and
  characterized with muons and electrons at the CERN East Area
  facility. The tests demonstrate that both the electromagnetic
  response and the sensitivity to minimum ionizing particles are
  retained after irradiation. Gain compensation can be achieved
  increasing the bias voltage well within the operation range of the
  SiPMs. The sensitivity to single photoelectrons is lost at
  $\sim 10^{10} \ \ncmsq$ due to the increase of the dark current.  }


\begin{document}
\maketitle
\flushbottom

\section{Introduction}
\label{sec:intro}

The possibility to use compact Silicon Photomultipliers (SiPMs)
embedded in the bulk of particle detectors is extremely
appealing. Direct coupling with scintillator or wavelength shifter (WLS)
fibers~\cite{calice,Berra:2016thx} remove the inefficiencies and dead
areas introduced by the light extraction toward conventional
photomultipliers (PMTs). Modern SiPMs are replacing PMTs in a vast
number of applications but the operation of these devices in radiation
harsh environments remains a
challenge~\cite{Musienko:2017znn,Cordelli:2017dgl,CentisVignali:2017zpz,Xu:2014vua,Garutti:2014jya}. Accelerator
neutrino physics applications  as developed by the ENUBET
Collaboration~\cite{enubet}   are in between low-dose environments
where damage due to non-ionizing particles is small ($< 10^{8} \
\ncmsq$)~\cite{Cattaneo:2017vgl} and high radiation environment as in
the forward region of high-luminosity colliders ($> 10^{13} \
\ncmsq$)~\cite{Heering:2016lmu}. In ENUBET, compact calorimeters are
employed to monitor lepton production in the decay tunnel of neutrino
beams at single particle level and to provide a 1\% measurement of the
neutrino flux at source. The monitored neutrino
beams~\cite{Longhin:2014yta} are narrow band beams where
particles in the tunnel are recorded only at large angles to identify the
decay product of kaons. As a consequence, particle rates and doses are
mostly due to hadron interactions.

\begin{figure}[!htb]
\centering
\includegraphics[width=0.7\columnwidth]{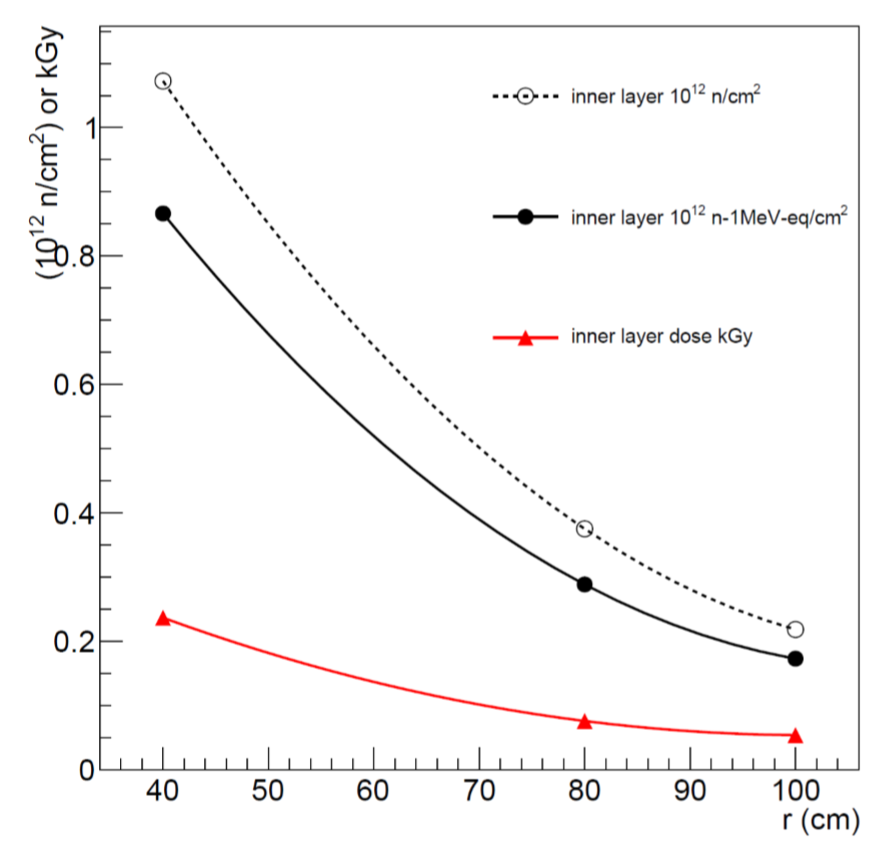}
\caption[]{\label{fig:dose} Ionizing doses (in kGy) and non-ionizing fluences (black dashed line: $\ncmsq$; black continuous line: 1 MeV-equivalent $\ncmsq$) as a function
of the distance between the axis of the ENUBET decay tunnel and the inner radius of the calorimeter.  }
\end{figure}

The integrated fluences to achieve a 1\% measurement of the $\nu_e$
and $\nu_\mu$ cross sections depend on the position of the calorimeter
with respect to the axis of the secondary beam (mostly pions and
kaons) at the entrance of the decay tunnel. Both ionizing radiation
doses and non-ionizing radiation fluencies are depicted in
Fig.~\ref{fig:dose} as a function of the distance between the beam
axis and the calorimeter. For ENUBET (1~m distance) the non-ionizing
fluence integrated during the lifetime of the experiment and scaled to
1~MeV equivalent neutrons is $1.8 \times 10^{11} \ \ncmsq$. The
ionizing dose is 0.06~kGy.

\begin{figure}[!htb]
\centering
\includegraphics[width=0.7\columnwidth]{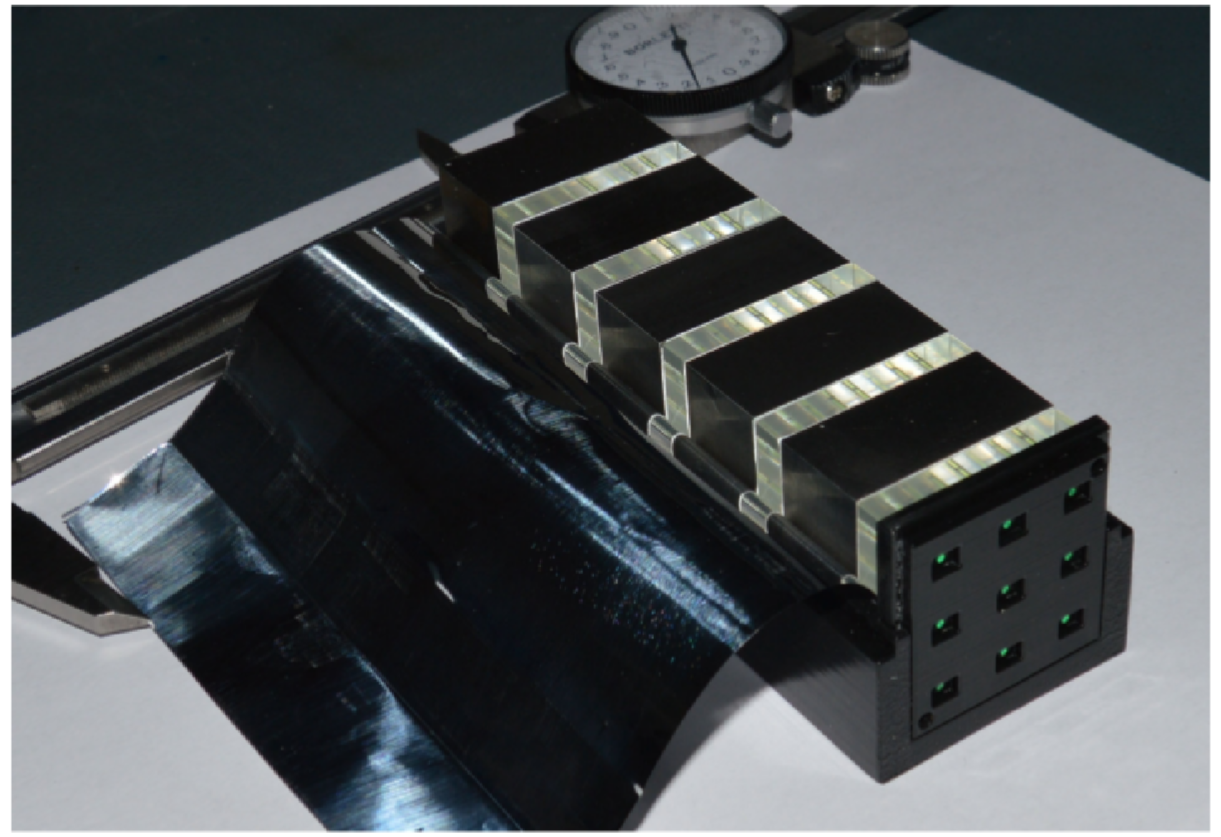}
\caption[]{\label{fig:UCM}   Picture of an ENUBET UCM employed for cosmic ray tests: the PCB hosting the SiPMs is
  not mounted and the plastic mask holding the fibers is visible in
  the back of the detector. The PCB (see Fig.~\ref{fig:PCB} below) is then mounted on the mask to couple the fibers to the photonsensors. }
\end{figure}

The ENUBET calorimeter is an assembly of Ultra Compact Modules (UCM -
Fig.~\ref{fig:UCM}). The basic module is an iron-plastic scintillator
device whose light is collected by WLS fibers
running perpendicularly to the absorber and converter plates
(``shashlik'' calorimeter~\cite{Fessler:1984wa,atoyan1992}). In a UCM,
every single fiber segment is directly connected to a
SiPM~\cite{Berra:2017rsi}. The array of SiPMs reading the UCM is
hosted on a PCB (Printed Circuit Board) holder that integrates both
the passive components and the signal routing toward the front-end
electronics. The calorimeters are assembled grouping arrays of UCMs,
whose size and thickness (in radiation lengths, $X_0$) are optimized
for the identification of positrons from \kethree. In ENUBET, the
use of compact calorimetric modules is a very effective solution but results into
exposing the SiPMs to fast neutrons produced by hadronic showers. The
SiPM technology of choice for ENUBET is RGB-HD: the  red-green-blue
(sensitivity) - high density  technology~\cite{acerbi2018} developed by FBK for small
($<25 \ \um$) pixel sensors with a sensitive area of the same size of
the ENUBET wavelength shifter fibers (1~mm$^2$). This setup has been
successfully tested with non-irradiated SiPMs and is described
in~\cite{Ballerini:2018hus}.  In 2017, the RGB-HD sensors were exposed
to fast neutrons at the Irradiation Test facility of INFN-LNL
(Sec.~\ref{sec:irradiation}). In Sec.~\ref{sec:analysis_LNL} we
describe the analysis of the dark current and noise waveform recorded
after irradiation at different fluences. The SiPMs were irradiated
in the same PCB boards used for the UCM and were tested at CERN to
establish the response to minimum ionizing particles (mips) and
electrons. The testbeams were performed at the CERN East Area facility
(T9 beamline) and are summarized in Sec.~\ref{sec:test_cern}.

\section{Irradiation tests at LNL}
\label{sec:irradiation}

\begin{figure}[!htb]
\centering
\includegraphics[width=\columnwidth]{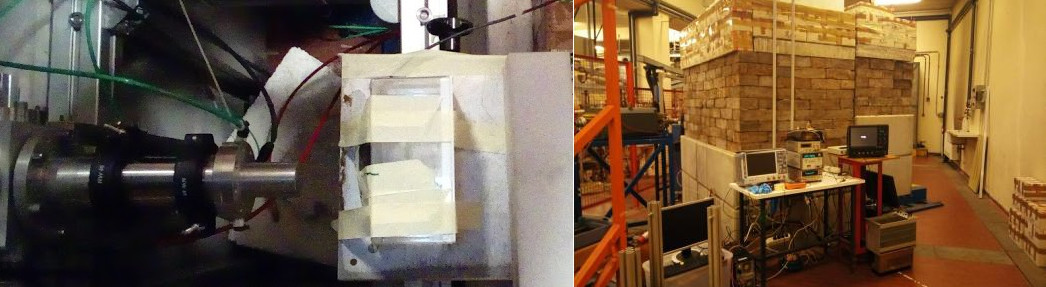}
\caption[]{\label{fig:test_area} (left) Top view of the irradiation test area at INFN-LNL: the sample holder (on the right) is located in front of the beam-pipe that hosts the Be target. (right) Experimental area and the setup to record the
dark current and the waveform of the SiPMs between two irradiation sessions. }
\end{figure}

The INFN-LNL (Laboratori Nazionali di Legnaro) provides a general
purpose irradiation facility based on the CN van der Graaf
accelerator~\cite{bisello2015}. The van der Graaf has a maximum voltage
of 7~MV and can accelerate protons and other light nuclei up to
5~$\mu$A currents. The tests were performed with a beam of protons
impinging on a thick Beryllium target. Neutrons are produced by
\ce{Be}(p,xn) reactions, namely \ce{^9Be}(p,n)\ce{^9B},
\ce{^9Be}(p,np) $2 \alpha$, \ce{^9Be}(p,np)\ce{^8Be} and
\ce{^9Be}(p,n$\alpha$)\ce{^5Li}. The irradiated sample is located
inside an experimental area with an external shield of concrete and an
inner shell of water as neutron moderator (see Fig.~\ref{fig:test_area}). 

\begin{figure}[!htb]
\centering
\includegraphics[width=0.9\columnwidth]{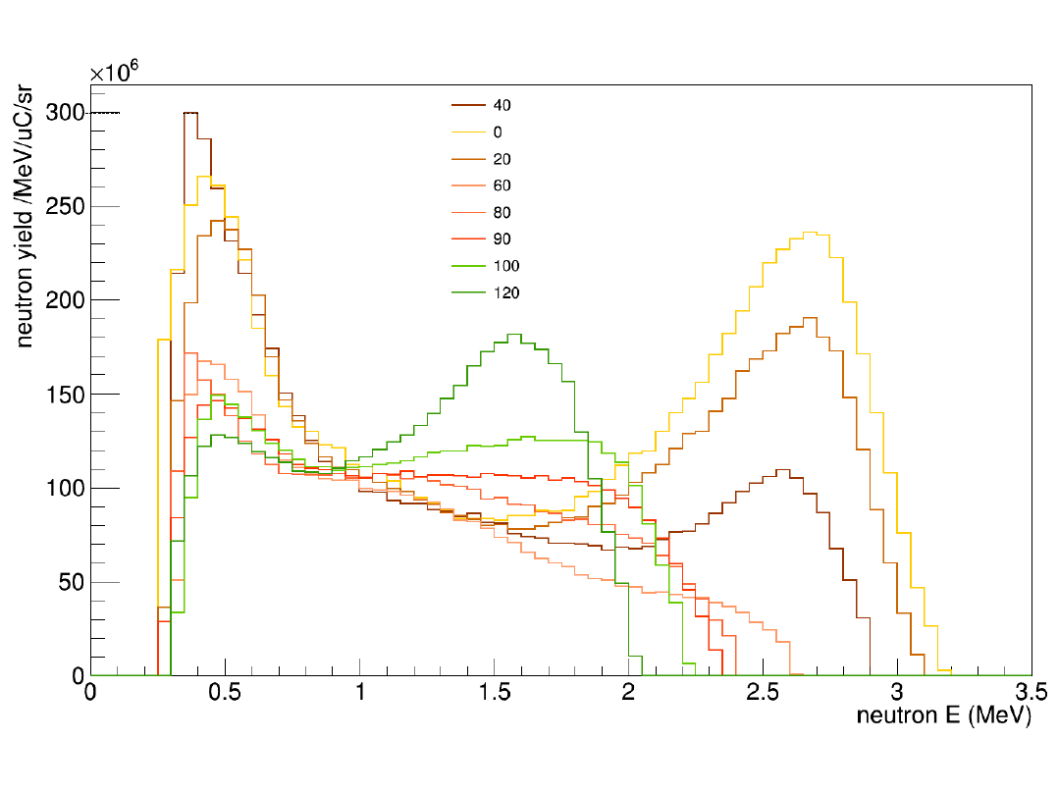}
\caption[]{\label{fig:neutron_spectra} Neutron yield as a function of energy (MeV) at different neutron emission angles (in degrees). Data from~\cite{agosteo_2011}.}
\end{figure}

A detailed assessment of
neutron yields for this beamline has been performed
in~\cite{agosteo_2011}. The neutron flux in the forward direction is peaked at about 0.5
and 2.7~MeV for 5~MeV protons impinging on the target. Fig.~\ref{fig:neutron_spectra} shows the neutron
yield per unit current (neutrons/MeV/$\mu$C/sr) at different angles (from $\theta=0^\circ$ up to $\theta=120^\circ$).
 The expected fluxes on the irradiated samples were evaluated
from~\cite{agosteo_2011} and from the real-time monitoring of the
proton current to the target performed with a current integrator.  Other
effects not included in~\cite{agosteo_2011} (neutron backscattering in
the shielding toward the sample) were estimated using FLUKA~2011~\cite{fluka_1,fluka_2} and
give a negligible contribution to the integrated fluence on the
sample.  

We irradiated three PCB boards used for the ENUBET UCMs. Each board
hosts 9 SiPMs and integrates both the passive components and the
signal routing toward the front-end electronics. The boards host 20,
15 and 12~$\um$ cell-pitch SiPMs, respectively. The SiPMs belonging to the
same UCM are connected in parallel and read out without amplification
through a 470~pF decoupling capacitor.  The PCB is equipped with a MCX
connector to read the sum of the current of the SiPMs and a miniature push-pull coaxial connector
(LEMO-00) for the bias. In addition, we assembled a test PCB with a
single 1 mm$^2$, 12~$\um$ pitch SiPM. Between two irradiation sessions, the
signal of the test PCB was connected to an Advansid trans-impedance amplifier (ASD-EP-EB-N ~\cite{advansid}) and the noise waveforms were recorded by a
Rohde \& Schwarz RTO 1024 oscilloscope to evaluate the single photoelectron
sensitivity. The current as a function of the overvoltage was measured
as voltage drop through a 10 k$\Omega$ resistor recorded by a Keithley
2700 multimeter and read out by a PC through a Keithley 7702 channel
multiplexer. The current of the SiPMs below the breakdown voltage was
$\mathcal{O}(10^{-9} \mathrm{A})$ for each SiPM and it was measured with a
Keithley 485 Picoammeter.  The temperature of the
sample during irradiation was measured with LM35 temperature sensors
recorded by an Arduino One microcontroller board.

\begin{figure}[!htb]
\centering
\includegraphics[width=0.6\columnwidth]{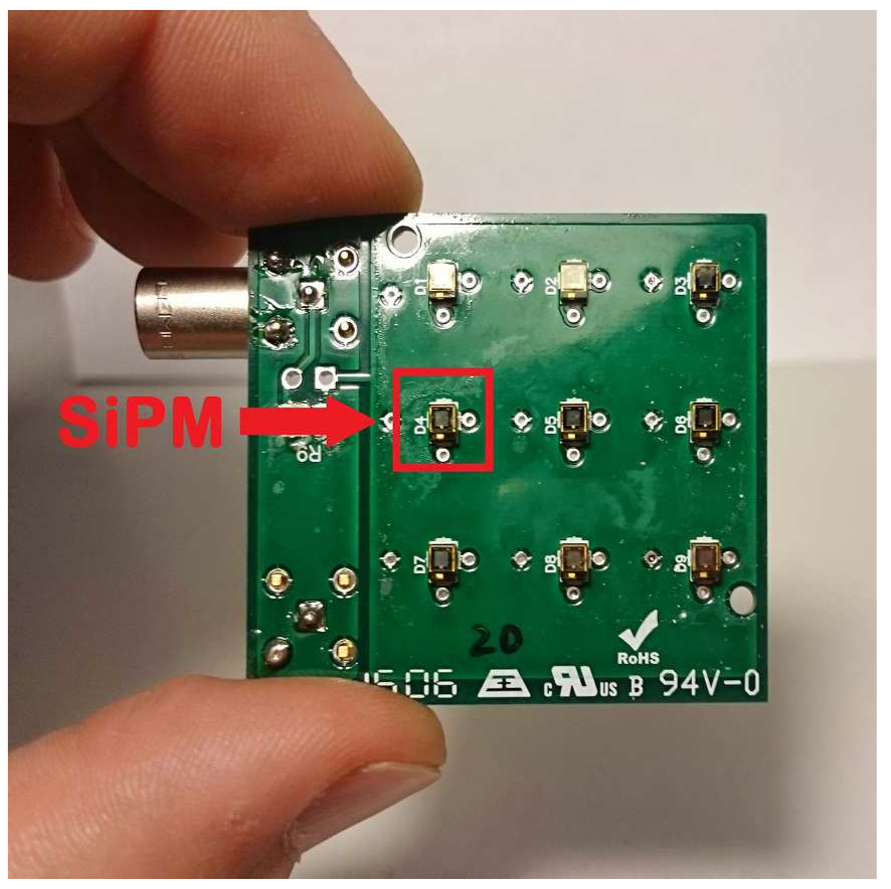}
\caption[]{\label{fig:PCB} One of the irradiated PCB with the SiPMs and the bias connector already installed.   }
\end{figure}

\section{Characterization of the irradiated SiPMs}
\label{sec:analysis_LNL}

The three PCBs hosting 9 SiPMs and the single-SiPM PCB were irradiated
from a minimum dose of $1.8\times 10^{8} \ \ncmsq$ up to $1.7\times
10^{11} \ \ncmsq$. During irradiation the SiPMs were not biased and
after each irradiation run, several current scans as a function of
voltage (I-V curve) were recorded. 
During the irradiation tests we employed two temperature probes with a
precision of 0.5$^\circ$C. The first one (``room temperature probe'')
was used to monitor room temperature (25~$^\circ$C during all the
irradiation period), and the second one (``sample probe'') was in thermal
contact with the irradiated sample. During irradiation, the SiPMs were
not biased and only the temperature of the sample was recorded. The
maximum increase of temperature during each irradiation run was
+10~$^\circ$C and the sample reached room temperature after 15-30
minutes.  The measurements reported below have been recorded when the
sample probe reached the value of the room-temperature probe.

\begin{figure}[!htb]
\centering
\includegraphics[width=0.8\columnwidth]{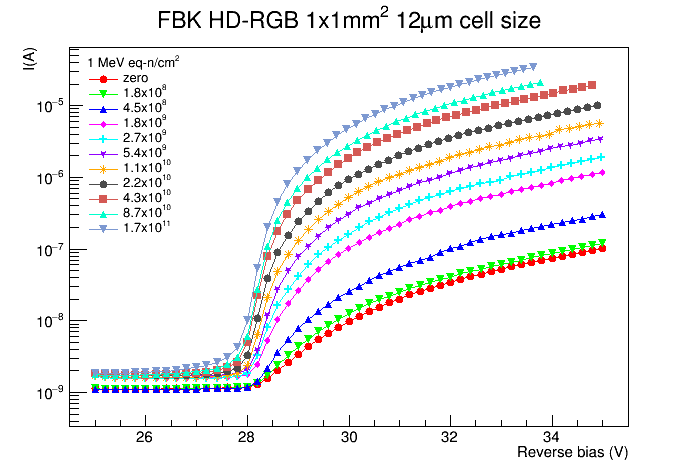}
\caption[]{\label{fig:ivcurve_single} I-V curve of the single-SiPM PCB.   }
\end{figure}

\begin{figure}[!htb]
\centering
\includegraphics[width=0.8\columnwidth]{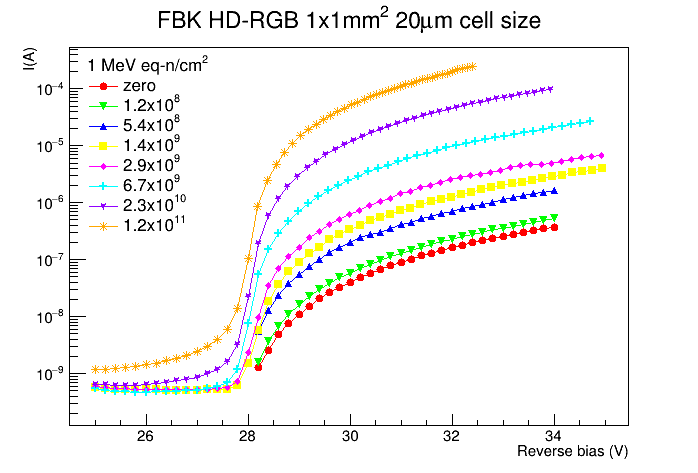}
\caption[]{\label{fig:ivcurve_20} I-V curve of the 20~$\um$ cell-pitch SiPMs. The value shown in the plot is normalized to one SiPM, i.e. the current from the PCB is divided by the number of SiPMs hosted in the board (9 SiPMs per board).   }
\end{figure}

\begin{figure}[!htb]
\centering
\includegraphics[width=0.8\columnwidth]{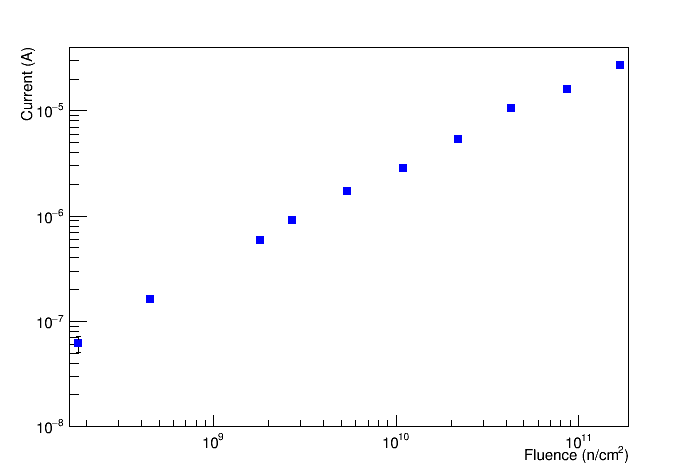}
\caption[]{\label{fig:dark_current} Dark current of the 12~$\um$ cell-pitch SiPMs (current of a 9 SiPM PCB divided by 9)
as a function of the neutron fluence. The bias of the SiPMs is 33~V, corresponding to +4.8~V overvoltage.  }
\end{figure}

Fig.~\ref{fig:ivcurve_single} shows the I-V
curves for the single-SiPM PCB (1 mm$^2$ area, 12~$\um$
cell-pitch). The curves were recorded with the sample at room temperature and, in the longest run (from
$8.7\times 10^{10}$ to $1.7\times 10^{11} \ \ncmsq$), thermal
equilibrium was reached 30 minutes after the stop of the proton beam. The corresponding
curves normalized to a single SiPM for the 20~$\um$ cell-pitch PCB are
shown in Fig.~\ref{fig:ivcurve_20}. The normalization is
performed dividing the value of the current from the PCB by the number of SiPMs hosted in the board (i.e. 9).
The current of all the 20~$\um$ pitch SiPMs was measured with a pico-ammeter before irradiation and is $\sim$ 0.5~nA at 27 V.     
The I-V curve of the 12 $\um$ cell-pitch PCB hosting 9 SiPMs is consistent with the corresponding single-SiPM PCB, i.e. the current is $9\times$ higher.  All the RGB-HD SiPMs show minor changes in the breakdown voltage. For the 12~$\um$ cell-pitch SiPM the breakdown voltage measured at the maximum of $I^{-1} dI/dV$ is 28.2~V for no irradiation and 28.0~V after an exposure of $1.7\times 10^{11} \ \ncmsq$. As expected, the dark current after breakdown increases by more than two orders of magnitude at a fluence of $\sim 10^{11} \ \ncmsq$. Fig.~\ref{fig:dark_current} shows the dark current normalized to 1~SiPM versus fluence at 33~V (+4.8~V overvoltage).

\begin{figure*}[t]
  \centering
  \subcaptionbox{}[.45\linewidth][c]{%
    \includegraphics[width=.45\linewidth]{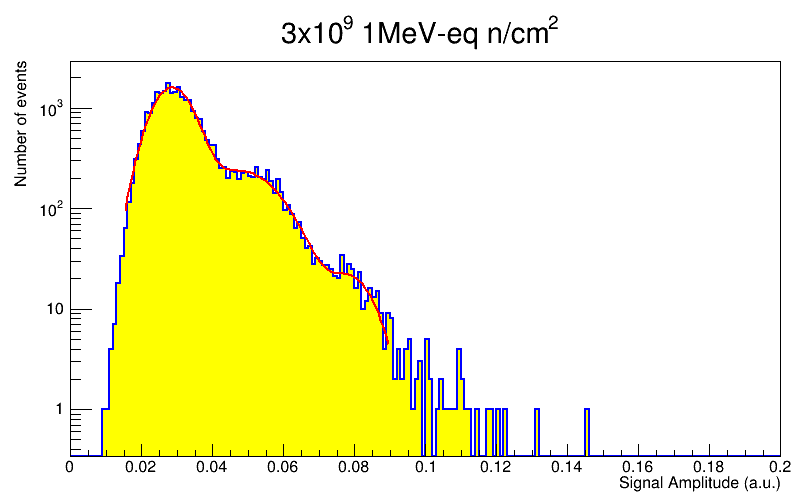}}\quad
  \subcaptionbox{}[.45\linewidth][c]{%
    \includegraphics[width=.45\linewidth]{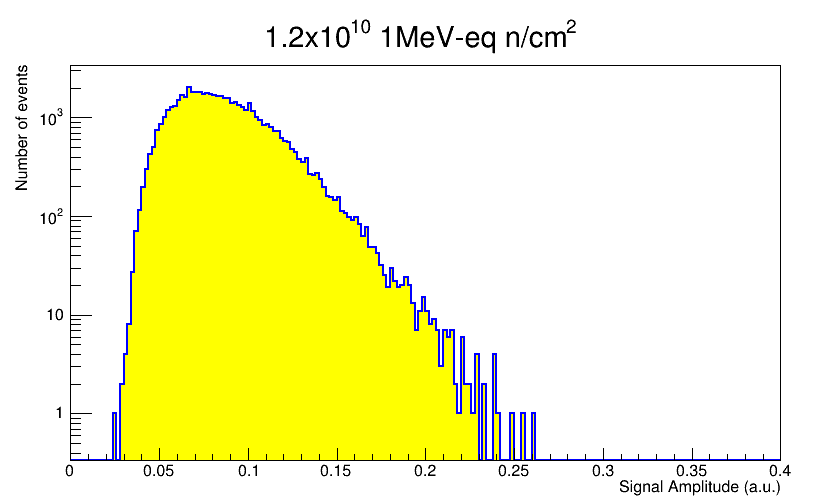}}\quad


  \caption{Signal peak distribution at  (a) $3 \times 10^{9}$ and (b) $1.2 \times 10^{10}$ $\ncmsq$  for the single-SiPM PCB (12~$\um$ cell-pitch, 1~mm$^2$ SiPM).}
\label{fig:waveform}
\end{figure*}

After each irradiation run, the signal terminals of the 12~$\um$
cell-pitch photosensor in the single-UCM PCB were connected to the
OUT2 of the ASD-EP-EB-N amplifier (transimpedance gain: 2500 for a 50
$\Omega$ load resistence) and the amplified waveform was recorded by the
Rohde \& Schwarz oscilloscope. The signal per photoelectron
(p.e.) was estimated operating the oscilloscope in self-triggering
mode and setting the threshold below the single photoelectron peak.
The waveform was sampled for 100~ns at 10~GS/s, i.e. we recorded 1000
samples every 100~ps for each triggered event. The distribution of the
signal peak is shown in Fig.~\ref{fig:waveform} at $3 \times
10^{9}$ (left) and $ 1.2 \times 10^{10}$ (right) $\ncmsq$.  The three fitted
peaks correspond to 1, 2 and 3~p.e. The sensitivity to single
photoelectron is lost at fluences larger than $3 \times 10^{9} \
\ncmsq $.

After irradiation the samples were stored at 25$\pm$1~$^\circ$C for
about three months before installing the boards on the UCM at CERN. In
this period, we expect the current to further decrease due to room
temperature self-annealing~\cite{Qiang:2012zh} and to reach a plateau
with a time constant of about 10 days. We have not studied the
behaviour of the dark current versus time during the storage period
and, hence, we do not report results on long-term self-anneling for
the HD-RGB. Before installation (see Sec.~\ref{sec:test_cern}),
however, we recorded the I-V curves and observed a current reduction
comparable to what reported in~\cite{Qiang:2012zh}.

\section{Tests on the T9 beamline}
\label{sec:test_cern}

\begin{figure}[hbtp]
\centering
\includegraphics[scale=0.3]{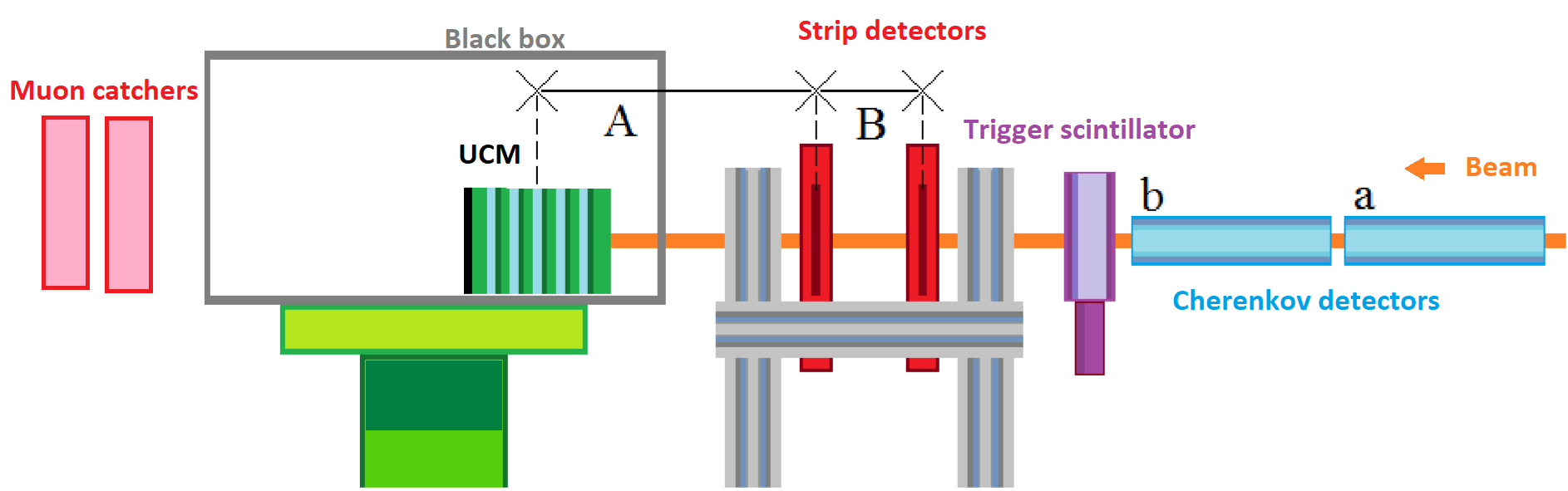} 
\caption{Schematics of the experimental setup at CERN (T9 beamline at the East Area)}
\label{fig:t9}
\end{figure}

The irradiated PCBs were tested on the T9 beamline~\cite{T9} of the
CERN East Area facility in October 2017 together with other ENUBET
prototypes. Figure~\ref{fig:t9} shows a schematics of the
instrumentation in the beam area. The particle beam is composed of
electrons, muons and pions and the momentum can be selected between 1
and 5~GeV, thus covering the whole range of interest for ENUBET. A
pair of threshold Cherenkov counters filled with CO$_2$ were used to
separate electrons from heavier particles.  The acquisition was
triggered by a 10$\times$10 cm$^2$ plastic scintillator located
downstream the Cherenkov detectors. We used a pair of silicon strip
detectors with a spatial resolution of $\sim$30~$\mu$m to track
charged particles down to the UCM. Two pads of plastic scintillator
(``muon catcher'') interleaved by a 20~cm thick iron shield were
positioned after the prototypes to identify muons or non-interacting
pions. The prototypes under test were positioned inside a light-tight
metallic box and mounted on a movable platform in front of the two
silicon strip detectors.  During the testbeam at CERN, the temperature
was monitored by probes placed inside the box containing the
calorimeter. In order to stabilize the temperature, we equipped the
box with a water-cooled chiller controlled by the probe in termal
contact with the UCM. The average temperature during the measurement
was 26$^\circ$C and the maximum variation was $\pm$1$^\circ$C.

Most of the tests were performed using the 15~$\um$ cell-pitch PCB irradiated up
to $1.2 \times 10^{11} \ \ncmsq$. The same UCMs were equipped with a
non-irradiated PCB identical to the irradiated one. The SiPMs of the
two PCBs belong to the same production batch and differences in
the breakdown voltage are less than 0.1~V.
The PCBs were tested with two ENUBET UCMs: 
\begin{itemize}
\item {\bf Prototype 16B} This is the first UCM prototype of ENUBET
  (Fig. \ref{fig:UCM}) and its design was employed for the calorimeter
  studied in~\cite{Ballerini:2018hus}. The UCM was assembled using five
  3$\times$3~cm$^2$ iron slabs with 1.5~cm thickness interleaved by
  five scintillator tiles (0.5 cm thickness). The slabs were drilled
  with a CNC machine: the distance between holes was 1~cm and the
  diameter of the holes was 1.2$\pm$0.2~mm. After drilling, the slabs were 
  zinc-plated to prevent oxidation. The 
  0.5~thick 3$\times$3 cm$^2$ scintillator tiles were machined and
  polished from EJ-200 \cite{ej200} plastic scintillator sheets and we
  inserted Tyvek foils between the scintillator and absorber tiles to
  increase the light collection efficiency. For 16B, we employed 1~mm diameter Kuraray
  Y11 fibers with an emission peak at 476 nm
  \cite{kuraray}. Laboratory tests on 16B show that a mip crossing the
  whole 16B UCM releases about 50 photoelectrons. The number of
  photoelectrons includes the efficiency of light production, light
  trasport to the photosensor and the photon detection efficiency (PDE) of
  the SiPM.

\item {\bf prototype 17UA} was built from injection molded
  scintillator tiles produced by Uniplast (Russia) \cite{uniplast} for ENUBET. In this
  prototype each tile is made by 3 extruded scintillator
  slabs (3$\times$3 cm$^2$, 4.5 mm thickness) for a total thickness of
  1.35~cm. The scintillator is polystyrene-based with 1.5$\%$
  paraterphenyl (PTP) and 0.01$\%$ POPOP. The surface of each tile was etched with
  a chemical agent to form a 30-100 $\mu$m layer that acts as
  a diffusive reflector in order to increase the light collection. The
  grooves in the mould form the holes for the 1-mm diameter WLS
  fibers. The UCM was assembled from five 1.5~cm iron slabs interleaved
  by five 1.35~cm scintillator tiles. As for 16B, we employed Kuraray Y11
  fibers since the light emission spectrum of the Uniplast
  scintillator is similar to EJ-200 and Y-11 are properly matched to
  both of them. Laboratory tests on 17UA performed in the same
  conditions as for 16A show that a mip crossing the whole 17UA UCM
  releases $\sim$85 photoelectrons.

\end{itemize}

\noindent
These prototypes were tested using both a 9-SiPM board that was not
irradiated at INFN-LNL and an irradiated board. The boards were
equipped with 15~$\mu$m cell-pitch SiPMs. The irradiated board hosts a SiPM
located in the top-right corner that was damaged 
before the irradiation and that was disconnected during the
measurements. This board hence reads 8 active SiPMs.

\begin{figure}[hbtp]
\centering
\includegraphics[scale=0.6]{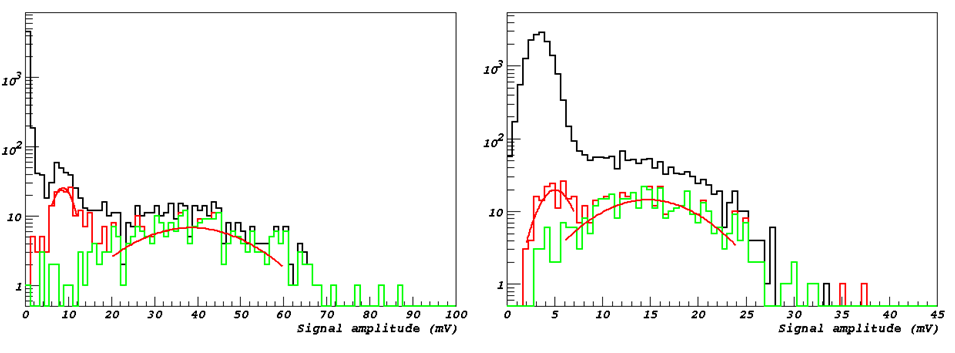} 
\caption{Signal in 16B of electrons (red) and mips (green) measured with the not-irradiated (left) and irradiated (right) PCB. The black line corresponds to all triggers and is dominated by dark noise (see text). The red lines show the Gaussian fit of the signal peak for electrons and mip-like particles. The SiPM overvoltage is +8.8~V and the beam momentum is 1 GeV.}
\label{fig:16B}
\end{figure}

\begin{figure}[hbtp]
\centering
\includegraphics[scale=0.4]{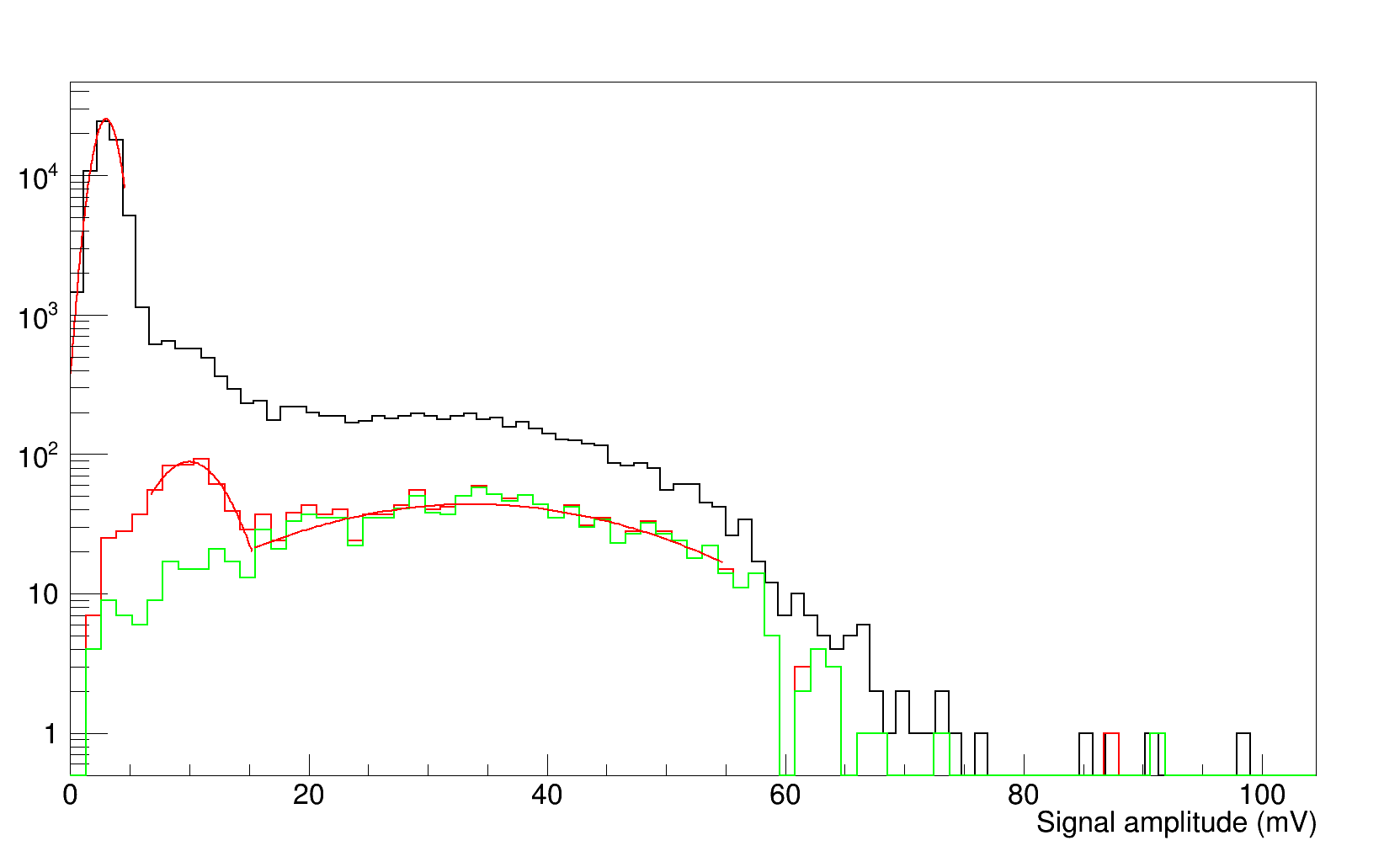} 
\caption{Electron (red) and mip (green) signal in 17UA with irradiated SiPM. The black line corresponds to all triggers and is dominated by dark noise (see text). The SiPM overvoltage is +8.8~V and the beam momentum is 1 GeV.}
\label{fig:17UA}
\end{figure}

Electrons were selected requiring a
signal in both Cherenkov counters.  Mip-like particles (muons or
non-interacting pions) correspond to events with no signal in the
Cherenkov counters and signal in the muon catcher.  The silicon
strip detectors are employed to select particles entering the front
face of the UCM in a $2\times2$~cm$^2$ fiducial area and crossing the
whole UCM.
Fig.~\ref{fig:16B} shows the signal response of 16B for
mips (green line) and electrons (red line). The left (right) plot
corresponds to the UCM with the non-irradiated (irradiated) SiPMs.
The black line show all signals triggered during the run and it is
dominated by dark counts.
 The loss of p.e. due to the missing SiPM was computed using a
GEANT4~\cite{Agostinelli:2002hh,Allison:2006ve,Allison:2016lfl}
optical simulation of the ENUBET UCMs.
The average p.e. loss due to the missing SiPM for the
particles selected in the fiducial area amounts to 10.6 $\pm$
0.1\%  (i.e. $\sim 1/9$).
A single UCM has a radiation length of 4.3~$X_0$ and covers
0.9~Moliere radii for particles entering at the center of the front
face. In the 1-3~GeV energy range, the first UCM hit by the electrons
is also the UCM with the maximum energy deposit although electrons are
only partially contained in the UCM and the width of the electron peak
is dominated by energy leakage. The signal of the first UCM for mip
and electrons (see Fig.~\ref{fig:16B} and~\ref{fig:17UA}) thus provides the detector
response in the whole dynamic range of interest for ENUBET.

The results of Fig.~\ref{fig:16B} demonstrate that a UCM collecting
50 photoelectrons per mip is not able to
separate a mip from the noise peak up to the maximum fluence
expected in ENUBET ($2 \times 10^{11} \ \ncmsq$) due to the increase of
the dark counts, even if  the electron peak remains well separated from noise. 
Preserving the sensitivity to mips for the entire duration of the run
is important in ENUBET for calibration purposes. The mip signal is employed to
monitor changes of the UCM response over the run - complementing the
LED monitoring system, - so that signal equalization over time can be
achieved increasing the overvoltage to compensate for amplitude
losses. In 16B, the number of photoelectrons per mip was limited to the poor
fiber-to-SiPM coupling, i.e. the mechanical tolerances in the plastic
mask that holds the fibers and couples them to the PCB. Since
the fiber diameter has the same size of the width of the SiPM
(1~mm), the photon collection efficiency is affected by misalignments between the
photosensors and fibers. 

\begin{figure}[hbtp]
\centering
\includegraphics[scale=0.40]{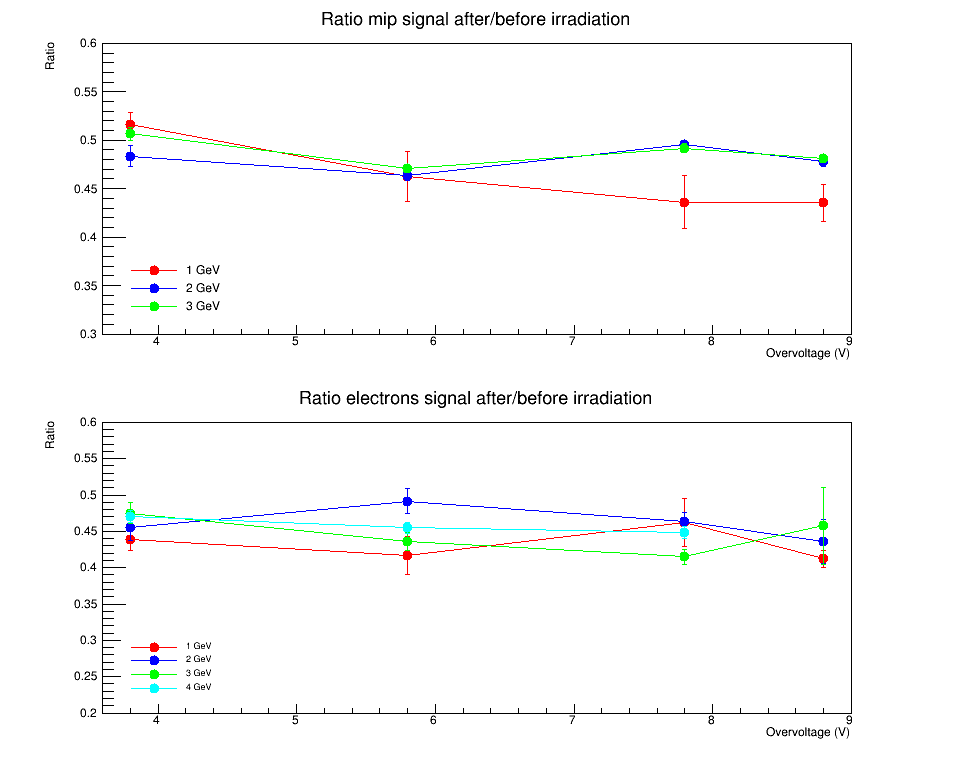} 
\caption{Ratio between the mip (upper plot) and electron (lower plot) peak signal amplitude after and before irradiation at different energies and overvoltage for 17UA.}
\label{fig:ratio}
\end{figure}

The 17UA prototype employs the same SiPM-to-fiber coupling scheme as
16B but has a larger scintillator thickness and the mip peak is
separated from the dark noise peak even after irradiation.  This is
demonstrated in Fig.~\ref{fig:17UA} for 1~GeV particles selected as in
16B.  The ratio between the mip peak after and before irradiation is
shown in Fig.~\ref{fig:ratio} (top plot) and is corrected for the
missing SiPM in the irradiated board. The bottom plot shows the
corresponding ratio for electrons at different energies. The overall
gain reduction is independent of the particle type, energy and
overvoltage within 5\%.  The electron and mip peak mean value ratio is
constant after irradiation and the integrated neutron fluence does not
affect the dynamic range of the photosensors. Hence, for the SiPMs
employed in this test (pixel size: 15~$\mu$m, fill factor: 62\%, pixel
density: 4444~pixels/mm$^2$) saturation effects of the signal due to
the reduction of the number of working pixels after irradiation are
not visible at $\mathcal{O}(10^{11} \ncmsq$). These effects may become
important at fluences of relevance for collider experiments where the
choice of the pixel size (smaller pixels to achieve the largest number
of cells per unit area) is a critical
parameter~\cite{Musienko:2017znn}.

Irradiation effects contribute to signal losses through a reduction of
the gain$\times$PDE and of the transparency of the epoxy employed for
the encapsulation of the SiPM. In this experimental setup, however,
non-irradiation effects due to board-to-board variations in the
SiPM-to-fiber coupling are sizable (20\%) and originate from the fact
that the width of the photosensor (1~mm) is the same as the diameter
of the WLS fiber. As a consequence, the coupling is sensitive to
mechanical displacement of the
photosensors~\cite{Ballerini:2018hus}. In fact, the signal loss of
Fig.~\ref{fig:ratio} represents a conservative estimate of the
change of response during the ENUBET data taking. A signal reduction
down to 50\% can be recovered by increasing the bias voltage to +5 V,
well within the operation range of the SiPM.

\section{Conclusions}
\label{sec:conclusions}

RGB-HD SiPMs produced by FBK and employed for calorimetric applications
at moderate neutron fluences were irradiated at LNL up to $10^{11} \
\ncmsq$ (1~MeV~eq). The dark current increases by two order of
magnitude when the neutron fluence goes from $10^{9}$ to $10^{11}
\ \ncmsq$. The single photoelectron sensitivity is lost at a fluence above
$3 \times 10^{9} \ \ncmsq$. Still, the photosensors can be safely operated in
calorimetric mode. An irradiated board of nine 1~mm$^2$ SiPMs with
15~$\mu m$ pixel size retains sensitivity to the mip if the number of
photoelectrons per mip is $\gtrsim 50$. At the maximum fluence ($1.2
\times 10^{11} \ncmsq$, 1~MeV~eq.), the relative response of the UCM
to electrons and mips is compatible with the response before
irradiation.

\acknowledgments

This project has received funding from the European Union's Horizon
2020 Research and Innovation programme under Grant Agreement
no. 654168 and no. 681647.  The authors gratefully acknowledge CERN
and the PS staff for successfully operating the East Experimental Area
and for continuous supports to the users.  We thank L. Gatignon,
M. Jeckel and H. Wilkens for help and suggestions during the data
taking on the PS-T9 beamline.  We are grateful to the INFN workshops
of Bologna, Milano Bicocca and Padova for the construction of detector
and to A.~Bau and G.~Pessina for support during the installation of
the SiPM.


\begin{thebibliography}{99}

\bibitem{calice}
C.~Adloff  et al. [CALICE Collaboration], 
\emph{Construction and Commissioning of the CALICE Analog Hadron Calorimeter Prototype,}
\jinst{5}{2010}{P05004}.

\bibitem{Berra:2016thx}
A.~Berra et al.,
\emph{A compact light readout system for longitudinally segmented shashlik calorimeters,}
\emph{Nucl.\ Instrum.\ Meth.\ A} {\bf 830} (2016) 345.


\bibitem{Musienko:2017znn}
  Y.~Musienko, A.~Heering, R.~Ruchti, M.~Wayne, Y.~Andreev, A.~Karneyeu and V.~Postoev,
\emph{Radiation damage in silicon photomultipliers exposed to neutron radiation,}
  JINST {\bf 12} (2017) no.07,  C07030.


\bibitem{Cordelli:2017dgl}
  M.~Cordelli, E.~Diociaiuti, R.~Donghia, A.~Ferrari, S.~Miscetti, S.~M\"uller and I.~Sarra,
\emph{Neutron irradiation test of Hamamatsu, SensL and AdvanSiD UV-extended SiPMs,}
  JINST {\bf 13} (2018) no.03,  T03005


\bibitem{CentisVignali:2017zpz}
  M.~Centis Vignali, E.~Garutti, R.~Klanner, D.~Lomidze and J.~Schwandt,
  \emph{Neutron irradiation effect on SiPMs up to $\Phi_{neq}$ = 5 $\times$ 10$^{14}$ cm$^{-2}$,}
  doi:10.1016/j.nima.2017.11.003
  arXiv:1709.04648 [physics.ins-det].

\bibitem{Xu:2014vua}
  C.~Xu, R.~Klanner, E.~Garutti and W.~L.~Hellweg,
\emph{Influence of X-ray Irradiation on the Properties of the Hamamatsu Silicon Photomultiplier S10362-11-050C,}
  Nucl.\ Instrum.\ Meth.\ A {\bf 762} (2014) 149

\bibitem{Garutti:2014jya}
  E.~Garutti, R.~Klanner, S.~Laurien, P.~Parygin, E.~Popova, M.~Ramilli and C.~Xu,
 \emph{Silicon Photomultiplier characterization and radiation damage investigation for high energy particle physics applications,}
  JINST {\bf 9} (2014) C03021.

\bibitem{enubet} 
A.~Berra et al. [ENUBET Collaboration],
\emph{Enabling precise measurements of flux in accelerator neutrino beams: the ENUBET project,}
CERN-SPSC-2016-036; SPSC-EOI-014.


\bibitem{Cattaneo:2017vgl}
  P.~W.~Cattaneo, T.~Cervi, A.~Menegolli, M.~Oddone, M.~Prata, M.~C.~Prata and M.~Rossella,
\emph{Radiation Hardness tests with neutron flux on different Silicon photomultiplier devices,}
  JINST {\bf 12} (2017) no.07,  C07012

\bibitem{Heering:2016lmu}
  A.~Heering, Y.~Musienko, R.~Ruchti, M.~Wayne, A.~Karneyeu and V.~Postoev,
\emph{Effects of very high radiation on SiPMs,}
  Nucl.\ Instrum.\ Meth.\ A {\bf 824} (2016) 111.

\bibitem{Longhin:2014yta}
A.~Longhin, L.~Ludovici and F.~Terranova, 
\emph{A novel technique for the measurement of the electron neutrino cross section,}
{\it Eur.\ Phys.\ J.\ C} {\bf 75} (2015) 155.

\bibitem{Fessler:1984wa}
H.~Fessler, P.~Freund, J.~Gebauer, K.~M.~Glas, K.~Pretzl, P.~Seyboth, J.~Seyerlein and J.~C.~Thevenin, 
\emph{A tower structured scintillator lead photon calorimeter using a novel fiber optics readout system,}
\nima{228}{1985}{303}.


\bibitem{atoyan1992}
G.~S.~Atoyan et al.,
\emph{Lead-scintillator electromagnetic calorimeter with wavelength shifting fiber readout,}
 \nima{320}{1992}{144}.


\bibitem{Berra:2017rsi}
  A.~Berra {\it et al.},
 \emph{Shashlik Calorimeters With Embedded SiPMs for Longitudinal Segmentation,}
  IEEE Trans.\ Nucl.\ Sci.\  {\bf 64} (2017) no.4,  1056.

\bibitem{acerbi2018}
F.~Acerbi, G.~Paternoster, A.~Gola, V.~Regazzoni, N.~Zorzi, C.~Piemonte, 
\emph{High-Density Silicon Photomultipliers: Performance and Linearity Evaluation for High Efficiency and Dynamic-Range Applications}
IEEE J.\ Quantum\ Elect.\ {\bf 54} (2018) 4700107. 

\bibitem{Ballerini:2018hus}
  G.~Ballerini {\it et al.},
 \emph{Testbeam performance of a shashlik calorimeter with fine-grained longitudinal segmentation,}
  JINST {\bf 13} (2018) P01028

\bibitem{bisello2015}
D.~Bisello  {\it et al.},
\emph{LNL irradiation facilities for radiation damage studies on electronic devices,} 
Nuovo Cimento {\bf C38} (2015) 189


\bibitem{agosteo_2011}
S. Agosteo {\it et al.}, 
\emph{Characterization of the energy distribution of neutrons generated by 5 MeV
protons on a thick Beryllium target at different emission angles,}
Appl. Rad. Isot. {\bf 69} (2011) 1664.

\bibitem{fluka_1} 
T.T. B\"ohlen, F. Cerutti, M.P.W. Chin, A. Fass\`o, A. Ferrari, P.G. Ortega, A. Mairani, P.R. Sala, G. Smirnov and V. Vlachoudis,
\emph{The FLUKA Code: Developments and Challenges for High Energy and Medical Applications,}
Nuclear Data Sheets {\bf 120}  (2014) 211

\bibitem{fluka_2} 
A. Ferrari, P.R. Sala, A. Fass\`o, and J. Ranft,
\emph{FLUKA: a multi-particle transport code,}
CERN-2005-10 (2005), INFN/TC\_05/11, SLAC-R-773. 

\bibitem{advansid} 
Advansid s.r.t., Via Sommarive 18, I-38123, Povo, Trento, Italy

\bibitem{Qiang:2012zh}
  Y.~Qiang, C.~Zorn, F.~Barbosa and E.~Smith,
  \emph{Radiation Hardness Tests of SiPMs for the JLab Hall D Barrel Calorimeter,}
  Nucl.\ Instrum.\ Meth.\ A {\bf 698} (2013) 234

\bibitem{T9} 
\texttt{http://sba.web.cern.ch/sba/BeamsAndAreas/East/East.htm}

\bibitem{ej200}{ELJEN Technology, 1300 W. Broadway, Sweetwater, TX 79556, USA}

\bibitem{kuraray} 
KURARAY CO., LTD., Ote Center Building,1-1-3, Otemachi, Chiyoda-ku, Tokyo 100-8115, Japan

\bibitem{uniplast} 
Uniplast OOO, Vladimir, st. Bolshaya Nizhegorodskaya, 77, Russia,

\bibitem{Agostinelli:2002hh}
  S.~Agostinelli et al. [GEANT4 Collaboration],
  \emph{GEANT4: A Simulation toolkit,}
  \emph{Nucl.\ Instrum.\ Meth.\ A} {\bf 506} (2003) 250.

\bibitem{Allison:2006ve} J.~Allison et al. [GEANT4 Collaboration], 
\emph{Geant4 developments and applications},
\itns{53}{2006}{270}.  

\bibitem{Allison:2016lfl} J.~Allison et al.,  \emph{Recent developments in Geant4,}
\emph{Nucl.\ Instrum.\ Meth.\ A} {\bf 835} (2016) 186.





\end{thebibliography}
\end{document}